\documentclass{article}
\usepackage{here}
\usepackage{epsfig}
\newcommand{\IGN}[1]{}
\begin{document}
\pagestyle{myheadings}
\title{\vspace*{-20mm}\mbox{\mbox{} \hspace*{24mm}\normalsize Published in: \underline{Lecture Notes in Computer Science {\bf 4131}, 710-717 (2006).}}\\[10mm]
Processing of information in synchroneously firing chains in networks of neurons}
\author{Jens Christian Claussen\\[6mm]\small 
Institut f\"ur Theoretische Physik und Astrophysik\\\small
Christian-Albrechts-Universit\"{a}t zu Kiel,\small
Leibnizstr.15,\\\small D-24098 Kiel, Germany\\
\small \texttt{http://www.theo-physik.uni-kiel.de/\~{ }claussen/} \\[-3mm] } \date{}
\maketitle        
\begin{abstract}
The Abeles model of cortical activity 
assumes that in absence of stimulation 
neural activity in zero order can be
described by a Poisson process.
Here the model is extended to describe 
information processing by synfire chains
within a network of activity uncorrelated to the
synfire chain.
A quantitative derivation of the transfer function
from this concept is given.
\end{abstract}
\setcounter{page}{710}
\pagestyle{myheadings} 
\markboth{J.\ C.\ Claussen \hfill Processing of information in synfire chains ~~~ } {J.\ C.\ Claussen \hfill Processing of information in synfire chains ~~~}

Two seminal concepts were introduced by Abeles
\cite{abeles81,abeles82,abeles91}:
A quantitative model for uncorrelated activity in the cortex in
absence of external stimulation, 
and the concept of the {\sl synfire chain},
a spatiotemporal pattern of synchroneous activity of
neurons being active in the same cortical task.

Synchroneous spiking, as a refinement of 
averaged firing rates,  has been used as an equivalent
mathematical basis for neural models 
\cite{gerstnerBC93,herrmannhertzbennett95}.
The experimental and theoretical aspects of
 synfire chains 
remain a field of active research
\cite{ikeda03,sougne01}
and also provide a conceptual basis for
neural computing architectures
\cite{wermter02}.
This paper analyzes
the extension to formulate processing and propagation
of information in such a network.

\section{The Abeles model of cortical activity}
\newcommand{\icp}{intracellular potential}
\newcommand{\icps}{\icp\,\,}
\hyphenation{intra-cellular}
The model of uncorrelated cortical activity given by Abeles
\cite{abeles81},
here referred to as Abeles Model, 
is a direct approach to understand why randomly firing by
self-excitation can be a stationary and robust firing mode in a neural
network. The underlying experiments are interpreted in the following way:
Even if the cortex is not excited by sensory input,
the neurons are firing randomly (Poisson process) and excite each other.
Obviously, this is to be interpreted as a ``ground state''
of the cortical network.
An interesting question is whether random firing is a stable mode of a
network or not. Because 99\% of the inputs to the cortex are coming from the
same or other cortical areas 
\cite{braitenberg},
we shall at first neglect the 1\% (sensory) input and therefore consider
a network with 100\% feedback.

\clearpage
\section{Definition of the Abeles model}
The defining assumptions to the Abeles Model are \cite{abeles81}:
\begin{itemize}
\item[(i)] Each postsynaptic potential has the shape of a falling exponential
(for $t\geq{}0)$:
\begin{eqnarray}\nonumber
 (+A) {\rm e}^{-t/ \tau} 
~~~~~~~~~~~~~
\mbox{for excitatory inputs,}
\end{eqnarray}
or,
\begin{eqnarray}\nonumber
(-A) {\rm e}^{-t/ \tau} 
~~~~~~~~~~~~~
\mbox{for inhibitory inputs.}
\end{eqnarray}
This assumption does not only include an idealization of the waveform, it
also includes that the values of synaptic strength $A$ and the time constants
$\tau$ are the same for all neurons.
 \\[-4mm]
\item[(ii)] All postsynaptic potentials sum up in a linear fashion giving the \icp;
the neuron generates a spike if the \icps reaches a threshold $T$.
 \\[-4mm]
\item[$\!$(iii)] All neurons are firing independently. This, however, is eqivalent to:
      No information is processed.
 \\[-4mm]
\item[$\!$(iv)] Each neuron has N synaptic inputs which can be excitatory or inhibitory
      in any proportion.
 \\[-4mm]
\item[(v)] The neurons fire at an average rate of $\lambda$ spikes per second.
\\[-4mm]
\end{itemize}

\section{The self-consistence equation for the average firing rate}
For high rates of inputs, the input spikes add up to nearly random fluctuations
 of the \icp; the probability density of the \icps therefore is Gaussian. 
This means: The firing rate of a cell is proportional to the probability for
the \icps to be above threshold:
\begin{equation}
\lambda = \frac{1}{\sigma} \cdot \frac{K}{\sqrt{2 \pi}} \cdot
             \int_{T}^{\infty} {\rm e}^{-\frac{x^{2}}{\sigma^{2}}} {\rm d}x
        = \frac{K}{\sqrt{2 \pi}} \cdot
             \int_{\frac{T}{\sigma}}^{\infty} {\rm e}^{-y^{2}} {\rm d}y
                                                    \label{eq:abeles1}
\end{equation}
where $K$ is an unknown constant and $\sigma^{2}$ is the variance of the
\icp, which can be calculated as follows:

Each postsynaptic potential contributes a variance of
\begin{eqnarray}\nonumber
\int_{0}^{\infty}((\pm A){\rm e}^{-t/ \tau})^2{\rm d}t =A^2 \frac{\tau}{2}.
\end{eqnarray}
Nota bene, excitatory and inhibitory connections here contribute equally.

The independent linear superposition of $N \cdot \lambda$ spikes (per second)
gives the total variance
\begin{eqnarray}
\sigma^2 = (N \lambda) \cdot (A^2  \frac{\tau}{2}),
\nonumber \end{eqnarray}
            \clearpage
 or \begin{eqnarray}
\frac{\sigma}{A} = \sqrt{N \lambda  \frac{\tau}{2}}. 
\label{eq:abeles2}
\end{eqnarray}
This means: For random firing at a {\em constant} average firing rate 
we have to satisfy a self-consistence-equation
\begin{equation}
\lambda = \frac{K}{\sqrt{2 \pi}} \cdot
             \int_{\frac{T}{
                    A  \sqrt{N \lambda  \frac{\tau}{2} }
                            }}^{\infty} {\rm e}^{-y^{2}} {\rm d}y
                            \label{eq:selbstkon}
\end{equation}
which still has $K$ and $T/A$ as free parameters to be fitted to the
experimental data.
Abeles' estimation for $T/\sigma$ is as follows:
If the neuron fires at a rate $\lambda$ and each spike is generated if the
membrane potential is approximately $1\mbox{ms}$ $(\approx 0.4\tau)$ above
threshold,
the probability of the \icps for being above threshold is approximately
$\lambda \cdot 1\mbox{ms} = 0.005,$
which is numerically equivalent to $T/\sigma$ being $ 2.58$.
Therefore only one parameter ($K$) is free, it can be evaluated by solving
equation~(\ref{eq:selbstkon}) for $K$.

The main results of the Abeles Model are 
quantitative estimations of network parameters from 
realistic neurophysiological properties.
Using $N = 20000, \lambda = 5 s^{-1} , \tau = 2.5 \mbox{ms,}$
$K = 1000 \mbox{s}^{-1},$  one obtains \cite{abeles82}:
\begin{itemize}
\item[(i)] $\sigma / A = \sqrt{125} \cong 11 $: The variance of the \icps is 11
times bigger than the amplitude of a single spike.
 \\
\item[(ii)] $T/A = T/\sigma \cdot \sigma / A \cong 2.58 \cdot 11 \cong 29 $:
Only $29$ {\em synchroneous} excitatory spikes will lift the membrane potential
to threshold. (This is a small value compared with $N\lambda = 100000$ spikes
that every cell receives per second.)
 \\
\item[$\!$(iii)] A {\em single} spike has no detectable effect on the output rate: The
firing rate increases from $5$ per second to $6.4$ per second, but relaxes
back to $5$ per second with the time con\-stant~$\tau$.
This causes only 0.003 extra output spikes.
 \\
\end{itemize}

\noindent
To conclude,
synaptic strength seems weak for detecting a single spike, but fairly strong
for detecting coincidence inputs. This is
a 
consequence of the
highly nonlinear error function, which determines by
(\ref{eq:abeles1}) the firing rate $\lambda$.
As analyzed in the appendix, below a critical firing rate
$\lambda_c$  random firing is unstable, so
that a certain level of activity is required to
transmit information.

\section{How can we describe processing of information?
-- Extension of the Abeles model}
\noindent
As one of the fundamental assumptions of the Abeles model is the
randomly firing of all neurons, which means that all spikes are
completely uncorrelated, it is {\em a priori} unable to describe
information transfer.

If the number of spikes carrying the information is much less
than the number of random spikes $(N\cdot\lambda\approx100000),$
the probability density of the \icps can be assumed to be approximately
Gaussian,
\hfill
 so that the mechanism 
\clearpage\noindent
is still the same:
The fluctuations converging to each neuron raise the \icps  to threshold.
Remarkably this condition does not explicitely restrict the correlated
activity of a {\em single}  neuron, so it can be involved constantly in
information processing.

How can we understand simple processing of information in a real network,
whose `ground state' is randomly firing at a rather low rate?
The concept given by Abeles is the {\em `synfire chain':}
Groups of synchroneously firing neurons are carrying the information;
their number must be sufficiently high (at least 10--20) to excite the
following neurons. 

A possible quantitative description of processing of information
within this concept is given by the model described in the remainder
of this paper.
The basic
 properties of the extended model  \cite{claussen89}
are defined as follows:
\begin{itemize}
\item[(i)] All input spikes --same as in the Abeles model-- are assumed to share
the common waveform of a falling exponential,  
\\
$x_i(t)=A_i {\rm e}^{-(t-t_0)/{\tau}} ({\rm for}\; t\geq 0),$
which may idealize the signal through the axon.
 \\[-4mm]
\item[(ii)] The synaptic strength, which was a constant $A$ in the Abeles model,
may be inhibitory $(A_i<0)$ or excitatory $(A_i>0),$
and is assumed to have different values for each neuron. In general, we
may assume the synaptic weights also to be time-dependent, so that
synaptic plasticity can be described. However, this time-dependence takes
place on a much larger time-scale than the spike dynamic.
 \\[-4mm]
\item[$\!$(iii)] All postsynaptic (episynaptic) potentials are assumed to sum up to the \icp:
\begin{equation}
I(t) = \sum_{i} A_i x_i(t).
\end{equation}
 \\[-4mm]
\item[$\!$(iv)] In addition to the Abeles Model we consider synchroneous and random
inputs seperately:
\begin{equation}
I(t) = \sum_{i(\mbox{\tiny sync})}   A_i x_i(t)
     + \sum_{i(\mbox{\tiny async})}  A_i x_i(t).
\end{equation}
As the number of randomly firing inputs is large, the difference in
synaptic strength will not disturb the Gaussian distribution, and we
can write for the second sum
\begin{equation}
\bar{A} \sum_{i(\mbox{\tiny async})}  x_i(t).
\end{equation}
This is the same property as in the Abeles Model, although the firing rate
may have a slightly different value.
\end{itemize}

For the synchroneous inputs, we now only consider one group of firing
neurons, so all these inputs have the same $t_0,$ so we can assume $t_0=0,$
and we have, writing $x_i(t)=X_i\cdot {\rm e}^{-t/{\tau}}:$
\begin{equation}
\sum_{i(\mbox{\tiny sync})}   A_i x_i(t)
=\sum_{i(\mbox{\tiny sync})}   A_i X_i \cdot {\rm e}^{-t/{\tau}}
= {\rm e}^{-(t-t_0)/{\tau}} \sum_{i(\mbox{\tiny sync})}   A_i X_i,
\end{equation}
\clearpage\noindent
where the $X_i$ are `digital' values (0 for no spike, 1 for a spike
correlated with the synfire chain).
Hence we can interprete the synchroneous inputs converging to the cell
as a time-dependent lowering of the potential threshold $T:$
\begin{equation}
I(t) - {\rm e}^{-t/{\tau}} \sum_{i(\mbox{\tiny sync})}   A_i X_i
= \bar{A}  \sum_{i(\mbox{\tiny async})}  x_i(t).
\end{equation}
Therefore, the firing rate is given by (all sums in the following text are
sums only over the synfire chain inputs):
\begin{equation}
\lambda(t) =\frac{K}{\sqrt{2\pi}}
\int_{\frac{  (T-\sum A_i x_i(t)) }{ \sigma  }}^{\infty}
{\rm e}^{-y^2} {\rm d}y,
\end{equation}
where $x_i(t)=X_i\cdot {\rm e}^{-t/{\tau}}.$ We shall write for the input sum:
\begin{equation}
{\cal X} := \sum A_i X_i.
\end{equation}

If we ask: What is the total number of extra spikes, generated by an input
${\cal X}\neq 0,$ 
i.e., 
$\Delta\lambda(t):= 
\lambda_{{\cal X}\neq 0}(t) -
\lambda_{{\cal X}  =  0}(t)$? 
--
We have to integrate the firing rate,
\begin{equation}
~~~~~
\int_0^{\infty} \!\!\! \Delta\lambda(t)  {\rm d}t
\!=\!\frac{K}{\sqrt{2\pi}} \!   \int_0^{\infty} \!\!\!   {\rm d}t \! \left[
  \int_{\frac{  (T-\sum A_i x_i(t)) }{ \sigma  }}^{\infty} {\rm d}y {\rm e}^{-y^2}
 \!\! - \!\!  \int_{\frac{  T }{ \sigma  }}^{\infty} \!\!\! {\rm d}y {\rm e}^{-y^2}
\!\right]\!,\!
\end{equation}
but this expression counts all extra
spikes from $t=0$ to $t=\infty.$
However, if the output shows too much time delay, it will not be correlated
to the synfire chain any more. As the time constant of the exponential
is $\tau,$ we only take into account the
outputs between $t=0$ and $t=\Delta t,$
where $\Delta t$ is a time constant which may have a similar or smaller
value than $\tau.$\\
So the average number of correlated output spikes
$\langle{\cal Y}\rangle$ to a given input ${\cal X}$ 
is given by:
\begin{equation}
\langle {\cal Y}({\cal X})\rangle  = \frac{K}{\sqrt{2\pi}}
\int_0^{\Delta t}  {\rm d}t
\int_{\frac{T-{\cal X}{\rm e}^{-t/{\tau}}}{\sigma}}^{\infty}  {\rm d}y {\rm e}^{-y^2}.
\label{eq:transferfn}
\end{equation}
Here we have {\em not} subtracted the accidental output spikes, for
their value is finite and rather small in this short time interval.
For ${\cal X}\rightarrow (-\infty)$ the average output  vanishes, which is
the limit of strong inhibitory inputs.
For ${\cal X}\rightarrow (+\infty),$ which is equivalent to strong
excitation, we obtain:
\begin{eqnarray}
\lim_{\mbox{\scriptsize$\cal   X$}\rightarrow\infty  }
\langle {\cal Y}({\cal X})\rangle  
&=& \frac{K}{\sqrt{2\pi}}
\int_0^{\Delta t} {\rm d}t \int_{-\infty}^{\infty}  {\rm d}y {\rm e}^{-y^2}
\nonumber
\\
&=& 
\frac{K}{\sqrt{2\pi}} \int_0^{\Delta t}  {\rm d}t \sqrt{\pi}
=\frac{K}{\sqrt{2}} \cdot \Delta t.   
\end{eqnarray}
If we choose our free parameter $\Delta t := \sqrt{2} / {K},$
the function $f(x) := \langle {\cal Y}({\cal X})\rangle$,
as defined by equation (\ref{eq:transferfn}),
is a function of sigmoid type and describes the probability
that an output spike is generated. For ${\cal X}=0$ we have
the probability of $0.005,$ which is the probability of accidental
output spikes.
\clearpage

We recognize this result as the McCulloch-and-Pitts
\cite{mccullochpitts} Neuron Model, but in a
fairly new light: Patterns of synchroneously firing neurons can be
transferred and processed in a quasi-digital manner even in a
randomly firing network, and the fluctuations are necessary
to understand the sigmoidal character of the response function.

\section{Conclusions and Outlook}%
Within the framework based on the activity model \cite{abeles82}
and the concept of synfire chains, 
it has been shown how processing of information can be described
quantitatively. 
 Considering correlated and uncorrelated neural activity seperately,
it is possible to describe information processing by synfire chains
through a network of (in ground state) randomly firing neurons
in a quantitative manner. 

The crudest idealizations concern the
waveform of the spikes.
 The stochastic description of the firing process and the
representation of `one bit' by more than one neuron are essential
 in the network for error-tolerance and the ability to generalization.

 For synchroneously firing groups of neurons the `quasi-digital'
McCulloch-and-Pitts neuron Model is valid; the fluctuations of the other
neurons determine the in\-put-out\-put characteristic to be sigmoidal.

 The extended model can be generalized in a straightforward manner
to describe also inhibitiory synapses 
and spatio-temporal aspects of real networks
by use of (on larger time-scales) time-dependent values $A_i(t)$ of
synaptic strength.
\noindent \normalsize \\[1.3ex] {Acknowledgment: The author gratefully acknowledges partial financial support by Deutsche Forschungsgemeinschaft (DFG) within SFB 654.}

\section*{Appendix: Stability analysis of the Abeles model}
We now investigate whether the fixed point satisfying the
self-consistence-equation (\ref{eq:selbstkon})
is stable or instable.
Although we do not know the exact dynamical properties of the network, we
can answer this question.
A small change in $\lambda$ will lead to a change in $\sigma,$ the variance
of the \icp, where $\sigma(\lambda(t),t)$ is given by
(\ref{eq:abeles2}).
The changed variance of the \icps will cause a change in the firing rate,
given by 
(\ref{eq:abeles1}).
However, this will need a
 certain delay $\Delta t,$
so that the stationary equation~(\ref{eq:abeles1}) has to be modified to
the iterative expression
\begin{equation}
\lambda(\sigma(t),t+\Delta t)
   = \frac{K}{\sqrt{2 \pi}} \cdot
   \int_{\frac{T}{\sigma(t)}}^{\infty} {\rm e}^{-y^{2}} {\rm d}y.
   \label{eq:abeles1it}
\end{equation}
Therefore we can approximate the real dynamics by the iteration
\begin{equation}
\lambda(\lambda(t),t+\Delta t) = \frac{K}{\sqrt{2\pi}} \cdot
\int_{\frac{T}{A \sqrt{N \lambda(t) \tau/2}}}^{\infty}
  {\rm e}^{-y^2} {\rm d}y               \label{eq:iteration}
\end{equation}
and we obtain the answer to an increase of $\lambda$ by the amount of
$\Delta \lambda:$
\begin{eqnarray}
\Delta\lambda(t+\Delta t)
&=&  \lambda(\sigma(\lambda(t)+\Delta\lambda),t+\Delta t) -
     \lambda(\sigma(\lambda(t)),t)   \nonumber \\
&=&  \frac{  \partial\lambda(\sigma(\lambda(t)),t)}{\partial\sigma(t)}
     \cdot  \frac{\partial\sigma(\lambda(t),t)}{\partial\lambda(t)}
     \cdot \Delta\lambda(t),  \nonumber
\end{eqnarray}
which means that every iteration stretches $\Delta\lambda$ by the factor
\begin{equation}
\alpha(\lambda) = \frac{\Delta\lambda(t+\Delta t)}{\Delta\lambda(t)}
= \frac{  \partial\lambda(\sigma(\lambda(t)),t)}{\partial\sigma(t)}
     \cdot  \frac{\partial\sigma(\lambda(t),t)}{\partial\lambda(t)}.
\end{equation}
Since

 $\frac{\partial}{\partial\lambda} (A\sqrt{N \lambda \tau/2}) =
       \frac{\sigma}{2\lambda},$
       and
\\
\begin{eqnarray}\nonumber
\frac{\partial}{\partial\sigma} (\frac{K}{\sqrt{2\pi}}
   \int_{\frac{T}{\sigma(t)}}^{\infty} {\rm e}^{-y^2} {\rm d}y )
   =
\frac{K}{\sqrt{2\pi}}(-{\rm e}^{-(T/\sigma)^2})\cdot(-\frac{T}{\sigma^2}),
\end{eqnarray}
we obtain
\vspace*{-.5ex}
\begin{equation}
\alpha = \frac{K}{\lambda} \frac{1}{2\sqrt{2\pi}} \frac{T}{\sigma}
         {\rm e}^{-(T/\sigma)^2}.   \label{eq:alpha1}
\end{equation}
For sufficiently small $\Delta\lambda$ the iteration values $\lambda_{i}$
are close to the start value $\lambda_{0},$ so that the Liapunov exponent
of the iteration is given by $L=\mbox{ln}|\alpha(\lambda_0)|.$
\hfill
Obviously 
   \clearpage\noindent
the fixed point, which is assumed to represent a `ground state'
of randomly firing, is a stable one if and only if the Liapunov exponent
is negative, which means that $|\alpha|<1.$
Using the experimental values given by Abeles
for the cortex of the cat,  $T/\sigma=2.58,$  $K=1000s^{-1},$
and $\lambda=5s^{-1},$ we obtain the Liapunov exponent $L=-2.02$ or
$\alpha=0.13,$ which is much less than $1.$ In this fixed point the
network gives strong damping to both fluctuations and external stimulus.
This includes also sufficient stability of the `Randomly Firing Mode':
Neither a fade-out nor a collective `explosion' of the firing
can be generated by small perturbations. To understand the effects of
 strong perturbations, we will take a 
 short view on the stability function
 $\alpha(T/\sigma)$.
Using equation~(\ref{eq:alpha1}), we have to remember that $\lambda/K$ is
a function of $T/\sigma,$ so that we can use the expression 
($x:=T/\sigma):$
\begin{equation}
\alpha(x) = \frac{ \frac{x}{2} {\rm e}^{-x^2} }{\int_{x}^{\infty}
            {\rm e}^{-y^2} {\rm d}y}.   \label{eq:alpha2}
\end{equation}
Two limiting cases can be considered:
For $x\rightarrow 0$,
which is the limes of very high firing rates,
$\alpha(x)$ is asymptotic to $x/\sqrt{\pi},$ so that $\alpha(x)$
decreases to zero. This expresses the damping of avalanche effects.
For $x\rightarrow \infty,$
which is the limes of very low firing rates, the integral is asymptotic to
$\frac{1}{2x}{\rm e}^{-x^2},$ therefore $\alpha(x)$ is asymptotic to $x^2$.
As $\alpha(x)$ is continuous, there must exist a critical firing rate
$\lambda_{c},$ where $\alpha(\lambda_{c})=1.$ It is the point where
the Liapunov exponent changes its sign.
If the firing rate is higher than $\lambda_{c},$ we still have damping,
same as in the ground state itself. If the firing rate is lower than the
critical value,  the cortical feedback amplifies any fluctuations of
the firing rate, so that the fluctuations lead to a fade-out of the network.
To conclude, if the firing rate is lower than a critical firing rate
$\lambda_{c},$ randomly firing cannot be a stable mode of a neural network.

\end{document}